\documentclass[10pt,twocolumn,showpacs,amsmath,amssymb,floatfix,superscriptaddress]{revtex4-2}
\usepackage{graphicx}
\usepackage{float}
\usepackage{dcolumn}
\usepackage{bm}
\usepackage{color}
\usepackage{txfonts}
\usepackage{microtype}
\usepackage{hyperref}
\usepackage{mathrsfs}
\usepackage{easyReview}
\usepackage{amsmath}
\usepackage{extarrows}
\usepackage{mathtools}
\usepackage{xcolor}
\usepackage{slashed}
\usepackage{amsmath}
\usepackage{ulem}

\setreviewson
\begin{document}

\title{Dynamics of relativistic vortex electrons in external laser fields}

\author{Mamutjan Ababekri}
\affiliation{Ministry of Education Key Laboratory for Nonequilibrium Synthesis and Modulation of Condensed Matter, Shaanxi Province Key Laboratory of Quantum Information and Quantum Optoelectronic Devices, School of Physics, Xi'an Jiaotong University, Xi'an 710049, China}
\author{Yu Wang}
\affiliation{Ministry of Education Key Laboratory for Nonequilibrium Synthesis and Modulation of Condensed Matter, Shaanxi Province Key Laboratory of Quantum Information and Quantum Optoelectronic Devices, School of Physics, Xi'an Jiaotong University, Xi'an 710049, China}
\author{Ren-Tong Guo}
\affiliation{Ministry of Education Key Laboratory for Nonequilibrium Synthesis and Modulation of Condensed Matter, Shaanxi Province Key Laboratory of Quantum Information and Quantum Optoelectronic Devices, School of Physics, Xi'an Jiaotong University, Xi'an 710049, China}
\author{Zhong-Peng Li}
\affiliation{Ministry of Education Key Laboratory for Nonequilibrium Synthesis and Modulation of Condensed Matter, Shaanxi Province Key Laboratory of Quantum Information and Quantum Optoelectronic Devices, School of Physics, Xi'an Jiaotong University, Xi'an 710049, China}
\author{Jian-Xing Li}\email{jianxing@xjtu.edu.cn}
\affiliation{Ministry of Education Key Laboratory for Nonequilibrium Synthesis and Modulation of Condensed Matter, Shaanxi Province Key Laboratory of Quantum Information and Quantum Optoelectronic Devices, School of Physics, Xi'an Jiaotong University, Xi'an 710049, China}
\date{\today}

\begin{abstract}
Investigating the interactions of vortex electrons with electromagnetic fields is crucial for advancing particle acceleration techniques, scattering theory in background fields, and developing novel electron beams for material diagnostics. In this work, we systematically study the dynamics of relativistic vortex electrons during their head-on collisions with linearly polarized (LP) and circularly polarized (CP) laser pulses, as well as their superposition. We develop a theoretical framework using Volkov-Bessel wave functions to describe the spatiotemporal characteristics of vortex electrons in these external fields. We show that the beam center of the vortex electron follows the classical trajectory of a point-charge electron while maintaining the transverse structure of both vortex eigenstates and superposition states. Specifically, CP laser pulses cause the beam center to rotate, while LP laser pulses induce a lateral shift. The combined effect of LP and CP laser pulses in a two-mode field results in a twisted spiral pattern. Our findings demonstrate the potential for versatile control of vortex electron beams using various laser modes, providing a foundation for future experimental and theoretical studies. This work serves as a benchmark reference for investigations into the manipulation of vortex electron beams using more realistic laser or other types of external fields.

\end{abstract}
\maketitle

\section{Introduction}
Vortex waves are non-plane-wave solutions to wave equations in cylindrical coordinates, characterized by helical wave fronts that carry intrinsic orbital angular momentum (OAM) along the propagation direction \cite{Padgett:17,bliokh2017theory,lloyd2017electron,knyazev2018beams}.  This was first identified in optics in 1992 when Allen and coworkers demonstrated that Laguerre-Gaussian laser modes carry intrinsic OAM \cite{allen1992orbital}. Intrinsic OAM of light, a new degree of freedom distinct from its spin angular momentum (SAM), gives rise to novel effects and opens up a vast array of applications, encompassing areas such as optical manipulation \cite{he1995direct,garces2003observation}, quantum optics \cite{mair2001entanglement,Wang:2015vrl}, and imaging \cite{swartzlander2001peering,swartzlander2008astronomical}, among others \cite{Padgett:17,knyazev2018beams}. Following Bliokh and coworkers' discovery \cite{Bliokh:2007ec}, vortex electrons were experimentally realized, in the 2010s, by reshaping the wave front of electron waves in scanning transmission electron microscope using spiral phase plates, diffraction gratings, and magnetic needles \cite{verbeeck2010production,Grillo:2014ksg,Beche:2013wua}. Similarly, vortex atoms and neutrons were also experimentally demonstrated in the following years \cite{Clark:2015rcq,Alon:2021,Sarenac_2018,sarenac2022experimental}. These vortex particles not only have emerged as a novel probe at the microscale, but also in the realm of high-energy physics, they hold great promise for advancing nuclear and particle physics research \cite{Ivanov:2019vxe,Ivanov:2022jzh,Lu:2023wrf,Xu:2024jlt,Lu:2024gha}. However, vortex particles are currently limited to low-energy regimes, and generating high-energy versions remains a primary goal in this field. Potential strategies for producing high-energy vortex leptons include direct generation through high-energy collision processes or accelerating lower-energy versions in external fields \cite{Ivanov:2022jzh}. Therefore, studying the manipulation of charged vortex particles using electromagnetic fields is critical for advancing these efforts.  \\

Relativistic vortex electrons can be described by Bessel wave solutions of the Dirac equation \cite{Bliokh:2011fi}. In contrast to vortex electromagnetic waves, vortex electrons manifest peculiar traits stemming from their charge $e$, rest mass $m_e$, and half-integer spin $s$, endowing them with qualities that set them apart while they still retain analogous vortex wave features \cite{bliokh2017theory,lloyd2017electron}. Vortex electrons display an inherent spin-orbit interaction (SOI) in free space, evidenced by the spin-dependent probability density of the electron beam \cite{Bliokh:2011fi}, and they also possess electromagnetic vortex fields produced by their spiral currents \cite{PhysRevLett.109.254801}.  Intense research has been devoted to other fundamental aspects of vortex electrons, including the mechanical properties \cite{PhysRevA.88.031802}, the relativistic effects ~\cite{PhysRevLett.118.114801,PhysRevLett.118.114802}, and radiation features \cite{Ivanov:2013eqa,Ivanov:2013bba,Kaminer:2014iia}. Furthermore, vortex electrons have been extensively studied within the contexts of vortex-particle-scattering picture \cite{VanBoxem:2013hpa,Seipt:2014bxa,Serbo:2015kia,Zaytsev:2016gqp,Sherwin:2018dah,Ivanov:2020kcy,Ivanov:2023hbp}. The effects of SOI and various vortex modes in the presence of external magnetic \cite{PhysRevX.2.041011,vanKruining:2017anw} and laser fields \cite{Hayrapetyan:2014faa,Bandyopadhyay:2015eri} have been explored.   Investigating the interaction of vortex electrons with external electromagnetic fields, particularly in the context of particle acceleration, has been approached using both semiclassical methods \cite{Silenko:2017fvf,Silenko:2018eed,Silenko:2019ziz} and wave-packet dynamics \cite{Baturin:2022utb,Sizykh:2023krl}. Vortex electrons can also be used to probe the magnetic structures of various materials and external fields \cite{PhysRevLett.108.074802,Guzzinati:2012mb,Grillo2017,PhysRevResearch.5.023192}. Consequently, a detailed investigation of vortex-electron interactions with laser fields promises to deepen our understanding of fundamental light-matter interactions and advance practical applications involving OAM-carrying electron beams.\\

To describe the interaction between a relativistic vortex electron and a plane-wave laser field, the Volkov solution to the Dirac equation in external fields can be employed to construct the Bessel vortex mode \cite{Karlovets:2012eu,Hayrapetyan:2014faa}.  Intense lasers play a crucial role in strong-field quantum electrodynamics  processes, owing to their rich parameter space that encompasses intensity, polarization, chirp, and various frequency modes \cite{DiPiazza:2011tq,Gonoskov:2021hwf,fedotov2022advances}.  In a linearly polarized (LP) laser pulse, the Volkov-Bessel state can be expanded in a series, with the $n$th mode characterized by total angular momentum (TAM) of $l+s+n$ and an OAM of $l+n$ under the paraxial approximation \cite{Hayrapetyan:2014faa}.  The coupling between OAM of electrons $l$ and the angular momentum (AM) of multiple laser photons $n$,  reflected by the term $n+l$, also appears in the context of vortex radiation and pair creation in circularly polarized (CP) lasers \cite{Karlovets:2022evc, Ababekri:2022mob, Guo:2023uyu, Ababekri:2024cyd}. In these investigations, the AM coupling proves vital for elucidating the mechanisms of AM transfer during the generation of vortex particles in strong laser fields. The oscillatory motion of vortex electrons due to the external plane-wave laser field is mentioned in the discussion of the analytical results obtained for monochromatic CP lasers \cite{Karlovets:2012eu}. A vortex state electron is found to undergo a circular motion in the transverse plane, which is related to the external OAM imparted by the laser \cite{Bu:2023fdj}.  
Furthermore, it has been demonstrated that LP lasers can induce a shift in the beam center of vortex electrons along the polarization direction of the external laser pulse, using parameters that are currently accessible in experimental settings \cite{Hayrapetyan:2014faa}. When the electron is propagating in an LP laser pulse, the vortex state of the electron is found to be destroyed by the highly unipolar laser pulse \cite{Aleksandrov:2022fmp}. Nevertheless, a systematic investigation into the manipulation of vortex electrons using various laser modes, including the specific instance of vortex electrons interacting with two-mode laser pulses, remains unexplored. \\

This work investigates the dynamics of relativistic vortex electrons in external laser fields by developing a theoretical formalism suitable for describing the interaction of vortex electrons with LP and CP pulses individually, as well as with the combined two-mode laser fields. We describe the vortex electron in the presence of external laser fields by constructing the Volkov-Bessel wave function, which is expressed as a sum over different vortex modes, explicitly showing the AM contributions from laser photons. We explore numerical results for the probability density of vortex electrons in CP and two-mode laser pulses. After discussing the relationship between the motion of the vortex electron, its momentum, and laser parameters, we highlight the advantages of two-mode laser fields in vortex-electron manipulation. The two-mode laser field offers richer possibilities and greater dimensions of control compared to LP or CP pulses used individually. \\

This paper is organized as follows. In Sec. \uppercase\expandafter{\romannumeral2}, we develop our theoretical formalism to describe the vortex electron's propagation in an external laser field. In Sec. \uppercase\expandafter{\romannumeral3}, we investigate the motion of the vortex electron in various external laser fields by presenting numerical results and discussing their physical meaning. In Sec. \uppercase\expandafter{\romannumeral4}, we summarize our findings and outline the implications of our work.\\

Throughout, natural units ($\hbar = c = 1$) are used, and the fine structure constant is $\alpha = \frac{e^2}{4\pi}$. The Feynman slash notation $\slashed{a} = a_\mu \gamma^\mu$ is employed.   

\section{Theoretical Formalism}
In this section, we develop our theoretical formalism for describing the propagation of vortex electrons in external laser fields. First, the construction of Bessel modes using plane-wave functions to describe the free vortex electron is reviewed for the sake of completeness. Following the same analogy, the vortex electron in an external laser field is described via the Volkov-Bessel state. We derive the four-current density as an observable quantity. Finally, we obtain the general result for the vortex electron propagating in two-mode laser fields, which is a superposition of LP and CP plane-wave fields with different central frequencies.\\

\subsection{Free vortex electron}
A free vortex electron can be described by constructing the Bessel-mode wave packet \cite{Ivanov:2011bv,Ivanov:2022jzh}, 
\begin{equation}
\psi^{\text{vor}}(x)=\int \frac{d^2\bm{p}_\perp}{(2\pi)^2}\,a_{l}(\bm{p}_\perp) \frac{u_{p}}{\sqrt{2\varepsilon}}e^{-ip\cdot x}\,,
\label{eq_free_vortex}
\end{equation} 
where $\frac{u_{p}}{\sqrt{2\varepsilon}}e^{-ip\cdot x}$ denotes the plane-wave component with four-momentum $p=(\varepsilon,\bm{p})$. 
The vortex amplitude $a_{l}(\bm{p}_\perp)$ is given by 
\begin{equation}
a_{l}(\bm{p}_\perp)=(-i)^{l}\,{e^{i\,l\,\phi_{\bm{p}}}}\,\sqrt{\frac{2\pi}{p_{\perp0}}}\delta(\vert\bm{p}_\perp\vert-p_{\perp0}),
\label{eq_vortex_amp}
\end{equation} 
where $l$ is the topological charge, $\phi_{\bm{p}}$ denotes the azimuth angle of the momentum vector, and $p_{\perp0}$ fixes the absolute value of the transverse  momentum. 
The Dirac bispinor can be expressed in the cylindrical coordinate as \cite{Hayrapetyan}
\begin{equation}
\begin{aligned}
u_p=\begin{pmatrix}\sqrt{\varepsilon+m_e}w^{(s)}\\ \sqrt{\varepsilon-m_e}\frac{p_z\sigma_z}{\vert\bm{p}\vert}w^{(s)} \end{pmatrix} + 
\begin{pmatrix}0 \\0 \\ \sqrt{\varepsilon-m_e}\frac{p_\perp}{\vert\bm{p}\vert} \begin{pmatrix} 0 \quad e^{-i\phi_{\bm{p}}} \\ e^{i\phi_{\bm{p}}} \quad 0	\end{pmatrix} w^{(s)} \end{pmatrix}\,,
\end{aligned}
\label{eq_Dirac_spinor_cylinder}
\end{equation} 
where spin eigenstates $w^{(s)}=(\alpha, \beta)^T$ of the operator $\sigma_z$ satisfy  $\sigma_z w^{(s)}=s w^{(s)}$, with $s=\pm\frac{1}{2}$. 
Moreover, considering the integral relation $\int \frac{d\phi_{\bm{p}}}{2\pi} (-i)^l e^{il\phi_{\bm{p}}}e^{i\xi\cos(\phi_{\bm{p}}-\phi_{\bm{r}})}
=e^{il\phi_{\bm{r}}}J_l(\xi)$, where $J_l$ is the Bessel function of the first kind, the integral in Eq.~\eqref{eq_free_vortex} can be evaluated by employing Eq.~\eqref{eq_Dirac_spinor_cylinder}. 
The final result reads \cite{Hayrapetyan}
\begin{equation}
\begin{aligned}
\psi^{\text{vor}}(x)&=\frac{e^{-i\Phi}}{\sqrt{2}}\sqrt{\frac{p_{\perp0}}{2\pi}} \\
&\times[\mathcal{A}_1 e^{il\phi_{\bm{r}}}J_l(\xi) + \mathcal{A}_2 e^{i(l+1)\phi_{\bm{r}}}J_{l+1}(\xi) + \mathcal{A}_3 e^{i(l-1)\phi_{\bm{r}}}J_{l-1}(\xi)],
\end{aligned}\label{eq_vortex_TAM}
\end{equation} 
where we define $\Phi=\varepsilon t - p_z z$ and $\xi=p_{\perp0} r$ and introduce the following notations:
\begin{equation}
\begin{aligned}
\mathcal{A}_1=\begin{pmatrix}
	\sqrt{1+\frac{m_e}{\varepsilon}} \begin{pmatrix} \alpha \\ \beta\end{pmatrix} \\
	\sqrt{1-\frac{m_e}{\varepsilon}} \cos\theta_{\bm{p}} \begin{pmatrix} \alpha \\ -\beta\end{pmatrix}
\end{pmatrix},
\mathcal{A}_2=\begin{pmatrix}
	0 \\ 0 \\ 0 \\ i\alpha\sqrt{\Delta}
\end{pmatrix},
\mathcal{A}_3=\begin{pmatrix}
	0 \\ 0 \\ -i\beta\sqrt{\Delta} \\ 0
\end{pmatrix}.
\end{aligned}
\end{equation} 
The polar angle of the momentum vector $\bm{p}$ is denoted by $\theta_{\bm{p}}$. The intrinsic SOI parameter, given by \(\Delta = \left(1 - \frac{m_e}{\varepsilon}\right)\sin^2\theta_{\bm{p}}\) \cite{Bliokh:2011fi}, vanishes in the paraxial limit ($\theta_{\bm{p}}\rightarrow0$) or the nonrelativistic limit ($\varepsilon\rightarrow m_e$). To explore the AM properties we introduce the following operators for SAM, OAM, and TAM along the $z$ axis, respectively:
\begin{equation}
\hat{\Sigma}_z=\frac{1}{2}\begin{pmatrix} \sigma_z \quad 0 \\ 0 \quad \sigma_z \end{pmatrix}\,,\quad \hat{L}_z=-i\frac{\partial}{\partial\phi_{\bm{r}}}\,, \quad \hat{J}_z= \hat{\Sigma}_z + \hat{L}_z\,.
\label{Operators}
\end{equation} 
The vortex wave function given by Eq.~\eqref{eq_vortex_TAM} satisfies the relation $\hat{J}_z\psi^{\text{vor}}(x) = (s + l)\psi^{\text{vor}}(x)$, indicating that the vortex electron is in an eigenstate of the TAM operator, carrying TAM eigenvalues $j = s + l$. Moreover, in the paraxial limit, the vortex state becomes an eigenstate of both SAM and OAM at the same time. 

\subsection{Vortex electron in the CP laser pulse}\label{sec_b}
\subsubsection{Volkov-Bessel electron}
In the presence of a plane-wave laser, a plane wave electron can be described using the Volkov wave function \cite{Berestetskii1982quantum},
\begin{equation}
\Psi_p(x)=(1+e\frac{\slashed{k}\slashed{A}(\varphi)}{2k\cdot p})\frac{u_{p}}{\sqrt{2\varepsilon}}e^{-ip\cdot x-i\frac{e}{k\cdot p}\int^\varphi d\varphi^\prime[p\cdot A(\varphi^\prime)-\frac{e}{2}A^2(\varphi^\prime)]}\,,
\label{Volkov}
\end{equation} 
where $A^\mu(\varphi)$ is the laser field, $\varphi=k\cdot x$ is the laser phase, $k=(\omega, 0, 0, -\omega)$ is the wave vector, and $\omega$ is the central frequency.  In the subsequent analysis, we consider a CP laser field, wherein the associated laser photons exhibit a SAM of one unit per photon. The vector potential of the laser field reads $A^\mu(\varphi) = a g(\varphi)(0, \cos\varphi, \sin\varphi,0)^T$, where $g(\varphi)$ is the pulse-shape function and $a$ is written as $a = a_0\frac{m_e}{e}$. The dynamic integral over the laser phase in Eq.~\eqref{Volkov} can be evaluated as follows:
\begin{equation}
\begin{aligned}
\frac{-e}{k\cdot p}\int^\varphi d\varphi^\prime p\cdot A(\varphi^\prime)=&\frac{e a p_\perp}{k\cdot p} \int^\varphi d\varphi^\prime g(\varphi^\prime) \cos(\phi_{\bm{p}}-\varphi') \\
\approx& \alpha_{p_\perp} g(\varphi)\sin(\phi_{\bm{p}}-\varphi)\,.
\end{aligned}
\label{SVEA}
\end{equation} 
Here, we introduce the notation $\alpha_{p_\perp}=-\frac{p_\perp m_e}{k\cdot p}a_0$ and have assumed the slowly varying envelope approximation (SVEA): $\partial_\varphi g(\varphi)\ll g(\varphi)$ \cite{seipt2016analytical}. 
Furthermore, after we introduce the Jacobi-Anger type series
\begin{equation}
e^{i \alpha_{p_\perp} g(\varphi)\sin(\phi_{\bm{p}}-\varphi)}=\sum_{n=-\infty}^{+\infty} J_n(\alpha_{p_\perp} g(\varphi)) e^{in(\phi_{\bm{p}}-\varphi)}\,,
\label{J_A}
\end{equation} 
the Volkov state for a CP laser pulse reads
\begin{equation}
\begin{aligned}
\Psi_p(x)=&(1+e\frac{\slashed{k}\slashed{A}(\varphi)}{2k\cdot p})\frac{u_{p}}{\sqrt{2\varepsilon}}e^{-ip\cdot x - i\beta_p\int d\varphi g^2(\varphi)}\\
&\times\sum_{n=-\infty}^{+\infty} J_n(\alpha_{p_\perp} g(\varphi)) e^{in(\phi_{\bm{p}}-\varphi)}\,,
\end{aligned}
\label{Volkov_CP}
\end{equation} 
where  $\beta_p=\frac{m^2_e}{2 k\cdot p}a^2_0$. Through a series expansion, we now explicitly obtain the azimuth angle $\phi_{\bm{p}}$-dependent phase factor $e^{in(\phi_{\bm{p}}-\varphi)}$. This factor originates from the laser-electron interaction and is crucial for understanding the OAM imparted by the absorption of multiple laser photons. \\

For a vortex electron in a laser field, we construct the Volkov-Bessel wave packet \cite{Hayrapetyan:2014faa}, given by $\Psi^{\text{vor}}(x)=\int \frac{d^2\bm{p}_\perp}{(2\pi)^2}\,a_{l}(\bm{p}_\perp)\Psi_p(x)$, where $\Psi_p(x)$ is the Volkov state in Eq.~\eqref{Volkov_CP}.
The wave function of the vortex electron in a CP laser field reads,
\begin{equation}
\begin{aligned}
\Psi^{\text{vor}}(x)=&(1+e\frac{\slashed{k}\slashed{A}(\varphi)}{2k\cdot p}) \sum_{n=-\infty}^{+\infty} i^n J_n(\alpha_{p_\perp} g(\varphi)) e^{-in\varphi} \tilde{\Psi}^{\text{vor}}_{n}(x)\,,
\end{aligned}
\label{volkov_Bessel}
\end{equation} 
where the $n$th mode function $\tilde{\Psi}^{\text{vor}}_{n}(x)$ reads
\begin{equation}
\begin{aligned}
\tilde{\Psi}^{\text{vor}}_{n}(x)&=\frac{e^{-i\tilde{\Phi}}}{\sqrt{2}}\sqrt{\frac{p_{\perp0}}{2\pi}}[\mathcal{A}_1 e^{i(l+n)\phi_{\bm{r}}}J_{l+n}(\xi) \\
& + \mathcal{A}_2 e^{i(l+n+1)\phi_{\bm{r}}}J_{l+n+1}(\xi) + \mathcal{A}_3 e^{i(l+n-1)\phi_{\bm{r}}}J_{l+n-1}(\xi)]\,,
\end{aligned}
\label{volkov_Bessel_n}
\end{equation} 
where the notation $\tilde{\Phi}=\varepsilon t - p_z z + \beta_p \int d\varphi g^2(\varphi)$ is introduced. Note that the Volkov-Bessel state is expressed as a sum over vortex modes, $\tilde{\Psi}_{n}(x)$, which satisfy $\hat{J}_z\tilde{\Psi}^{\text{vor}}_{n}(x) = (s + l+n)\tilde{\Psi}^{\text{vor}}_{n}(x)$. These modes carry TAM $j_n=s+l+n$. When the background laser field is turned off, $a_0\rightarrow0$, one obtains the wave function of the free vortex electron, as $\Psi^{\text{vor}}(x)=\tilde{\Psi}^{\text{vor}}_{n=0}(x)=\psi^{\text{vor}}(x)$.  \\

\subsubsection{Probability density}
As an important observable related to the Volkov-Bessel state given by Eq.~\eqref{volkov_Bessel}, the four-current density can be defined as
\begin{equation}
\mathcal{J}^\mu=(\rho_{l,s},\bm{j}_{l,s})=\bar{\Psi}^{\text{vor}}\gamma^\mu \Psi^{\text{vor}},
\label{four_current}
\end{equation}
where $\bar{\Psi}^{\text{vor}}=\Psi^{\text{vor}\,\dagger}\gamma^0$.  The probability density ($\rho_{l,s}=\mathcal{J}^0$) reads
\begin{equation}
	\begin{aligned}
		\rho_{l,s}=&\sum_{n,n^\prime} J_{n^\prime}( \alpha_{p_\perp} g(\varphi) )J_n(\alpha_{p_\perp}g(\varphi)) \Biggl(
		\cos \bigl[ (n-n^\prime) \bigl(\tilde{\phi}_{\bm{r}}+\frac{\pi}{2}\bigr) \bigr] \\
		&\times\Biggl[(1+\frac{\tilde{\delta}^2}{2})\rho_{lsnn^\prime}(\xi)+\frac{\tilde{\delta}^2}{2}\frac{p_z}{\varepsilon}J_{l+n^\prime}(\xi)J_{l+n}(\xi)\Biggr] \\
		&+\sin \bigl[ 2s (n-n^\prime) \bigl(\tilde{\phi}_{\bm{r}}+\frac{\pi}{2}\bigr) + \tilde{\phi}_{\bm{r}} \bigr] 
		\frac{\tilde{\delta} \, p_\perp}{\varepsilon}J_{l+n^\prime}(\xi)J_{l+n+2s}(\xi) \Biggr),
	\end{aligned}\label{rho}
\end{equation}
where $\tilde{\phi}_{\bm{r}}=\phi_{\bm{r}}-\varphi$ and $\tilde{\delta}=\frac{\omega m_e a_0}{k\cdot p}g(\varphi)$.  The following definition is introduced:
\begin{equation}
\rho_{lsnn^\prime}(\xi)=(1-\frac{\Delta}{2})J_{l+n^\prime}(\xi)J_{l+n}(\xi)+\frac{\Delta}{2}J_{l+n^\prime+2s}(\xi)J_{l+n+2s}(\xi)\,.
\label{rho00}
\end{equation}
Vector components of the current density in Eq.~\eqref{four_current} are given in the Appendix A. An overall factor of \(\frac{P_{\perp0}}{2\pi}\) in front of the four-current densities is omitted throughout this paper, as it does not alter the physical results in this work.\\  

Compared to the results for the LP laser pulse in Ref.~\cite{Hayrapetyan}, $\rho_{l,s}$ in Eq.~\eqref{rho} differs slightly. Here, the probability density features a ``mirror reflection'' for a new variable, $\tilde{\phi}_{\bm{r}}\rightarrow\tilde{\phi}_{\bm{r}}-\pi$, instead of the azimuth angle $\phi_{\bm{r}}\rightarrow\phi_{\bm{r}}-\pi$. Hence, the probability density's symmetric axis varies as a function of laser phase $\varphi$.  To focus on the dynamics of the vortex electron while neglecting radiation effects, we consider a relatively weak laser ($\tilde{\delta} \ll 1$) and an electron energy on the order of $\varepsilon \sim \text{MeV}$. In such cases, the probability density can be approximated as follows:
\begin{equation}
	\begin{aligned}
		\rho_{l,s} &\approx \sum_{n,n^\prime} J_{n^\prime}(\tilde{\alpha}_{p_\perp}) J_n(\tilde{\alpha}_{p_\perp}) \cos \bigl[ \Delta n \bigl(\tilde{\phi}_{\bm{r}} + \frac{\pi}{2} \bigr) \bigr] \rho_{lsnn^\prime}(\xi),
	\end{aligned}\label{rho_simp}
\end{equation}
where we introduce the definitions $\tilde{\alpha}_{p_\perp} = \alpha_{p_\perp} g(\varphi)$ and $\Delta n = n - n'$. From this point on, we will present results for $\tilde{\delta} \ll 1$ and defer the complete expression to the Appendix B.

When the electron is in a superposition state of the two vortex eigenmodes $a_{l_1}(\bm{p}_\perp)$ and $a_{l_2}(\bm{p}_\perp)$, which are defined as in Eq.~\eqref{eq_vortex_amp}, the vortex amplitude is given  by: $a_{l_1,l_2}(\bm{p}_\perp)=\frac{1}{\sqrt{2}}\left[a_{l_1}(\bm{p}_\perp)+e^{i\delta^\ast}a_{l_2}(\bm{p}_\perp)\right]$. The relative phase $\delta^\ast$ can take an arbitrary value.  In this case, the probability density reads
\begin{equation}
	\begin{aligned}
		\rho_{l_1l_2s} &\approx \sum_{n,n'} \frac{1}{2} J_{n'}(\tilde{\alpha}_{p_\perp}) J_{n}(\tilde{\alpha}_{p_\perp})  \\
		&\times  \Biggl\{
		\cos[\Delta n(\tilde{\phi}_{\bm{r}} + \frac{\pi}{2})] \Biggl[
		(1 - \frac{\Delta}{2}) (J_{l_1+n'} J_{l_1+n} + J_{l_2+n'} J_{l_2+n}) \\
		&\qquad + \frac{\Delta}{2} (J_{l_1+n'+2s} J_{l_1+n+2s} + J_{l_2+n'+2s} J_{l_2+n+2s})
		\Biggr] \\
		&\quad + 2 \cos[\Delta n(\tilde{\phi}_{\bm{r}} + \frac{\pi}{2}) + \Delta l \phi_{\bm{r}} + \delta^\ast] \Biggl[
		(1 - \frac{\Delta}{2}) J_{l_1+n'} J_{l_2+n} \\
		&\qquad + \frac{\Delta}{2} J_{l_1+n'+2s} J_{l_2+n+2s}
		\Biggr]
		\Biggr\}.
	\end{aligned}
	\label{sup_rho}
\end{equation}
Here, we omit the argument $\xi=p_{\perp0} r$ for the Bessel functions that appears inside the parentheses, starting from the second line. If we set $l_1,l_2=l$ and $\delta^\ast=0$ in Eq.~\eqref{sup_rho}, we recover the result in Eq.~\eqref{rho_simp}.  \\

\subsection{Vortex electron in the two-mode laser pulse}
Now we consider the scenario of a vortex electron colliding with a two-mode laser field, $A_{\text{two}} = A_1(\varphi_1) + A_2(\varphi_2)$, where $A_1$ is an LP laser pulse and $A_2$ is a CP laser pulse. The vector potential is given as,
\begin{equation}
A^\mu_{\text{two}} =a_1g_1(\varphi_1)\begin{pmatrix} 0 \\ 0\\ \sin\varphi_1 \\ 0 \end{pmatrix}+a_2g_2(\varphi_2)\begin{pmatrix} 0 \\ \cos\varphi_2\\ \sin\varphi_2 \\ 0 \end{pmatrix}\,.
\label{two_field}
\end{equation}
The laser frequencies satisfy $\omega_2 = \nu\omega_1$; hence, we have $k_2=\nu k_1$ and $ \varphi_2 = \nu\varphi_1 $. The laser pulse shapes are given by $g_1(\varphi_1)$ and $g_2(\varphi_2)$. 
The following treatment is employed to carry out the dynamic integral over the laser phase:
\begin{equation}
	\begin{aligned}
		\frac{-e}{k\cdot p} &\int^{\varphi_1}  d\varphi^\prime_1 p [A_1(\varphi^\prime_1) + A_2(\nu\varphi^\prime_1)]  \\
		&\approx\alpha_{1,p_\perp} g_1(\varphi_1) \cos\varphi_1 \sin\phi_{\bm{p}} + \alpha_{2,p_\perp} g_2(\nu\varphi_1) \sin(\phi_{\bm{p}}\nu\varphi_1),
	\end{aligned}
	\label{integral}
\end{equation}
where we have assumed SVEA and introduced $\alpha_{1,p_\perp}=-\frac{p_\perp m_e}{k_1\cdot p}{a}_{1,0}$ and $\alpha_{2,p_\perp}=-\frac{p_\perp m_e}{k_2\cdot p}{a}_{2,0}$. 
After introducing $\tilde{\alpha}_{1,p_\perp}=\alpha_{1,p_\perp}g_1(\varphi_1)\cos\varphi_1$ and $\tilde{\alpha}_{2,p_\perp}=\alpha_{2,p_\perp}g_2(\varphi_2)$, we can  write down the following relation:  
\begin{equation}
\begin{aligned}
&e^{i\tilde{\alpha}_{1,p_\perp}\sin\phi_{\bm{p}}+i\tilde{\alpha}_{2,p_\perp}\sin(\phi_{\bm{p}}-\varphi_2)}\\
&\equiv\qquad\sum_{n_1,n_2}J_{n_1}(\tilde{\alpha}_{1,p_\perp})J_{n_2}(\tilde{\alpha}_{2,p_\perp})e^{in_1\phi_{\bm{p}}}e^{in_2(\phi_{\bm{p}}-\varphi_2)}\,.
\end{aligned}
\label{J_A_two}
\end{equation} 
The Volkov-Bessel state in the two-mode laser field $A_{\text{two}}$ reads
\begin{equation}
	\begin{aligned}
		\Psi_{\text{two}}^{\text{vor}}(x)=&(1+e\frac{\slashed{k}\slashed{A}_{\text{two}}(\varphi_1)}{2k\cdot p}) \\
		&\times\sum_{n_1,n_2}i^{n_1+n_2} J_{n_1}(\tilde{\alpha}_{1,p_\perp})J_{n_2}(\tilde{\alpha}_{2,p_\perp}) e^{-i n_2\varphi_2} \tilde{\Psi}_{n_1,n_2}^{\text{vor}}(x)\,,
	\end{aligned}
	\label{two_volkov_Bessel}
\end{equation} 
in which, the mode function $\tilde{\Psi}_{n_1,n_2}^{\text{vor}}(x)$ reads
\begin{equation}
	\begin{aligned}
		\tilde{\Psi}^{\text{vor}}_{n_1,n_2}(x)&=\frac{e^{-i\tilde{\Phi}_\text{two} }}{\sqrt{2}}\sqrt{\frac{p_{\perp0}}{2\pi}}[\mathcal{A}_1 e^{i(l+n_1+n_2)\phi_{\bm{r}}}J_{l+n_1+n_2}(\xi) \\
		&\qquad + \mathcal{A}_2 e^{i(l+n_1+n_2+1)\phi_{\bm{r}}}J_{l+n_1+n_2+1}(\xi)  \\
		&\qquad + \mathcal{A}_3 e^{i(l+n_1+n_2-1)\phi_{\bm{r}}}J_{l+n_1+n_2-1}(\xi)]\,,
	\end{aligned}
	\label{two_volkov_Bessel_n}
\end{equation} 
where the notation $\tilde{\Phi}_\text{two}=\varepsilon t - p_z z + \Phi_\text{two}(\varphi_1)$ is introduced with the definition $\Phi_\text{two}(\varphi_1)=-\frac{e^2}{2k_1\cdot p}\int d\varphi'_1 [a_1^2 g_1^2(\varphi_1)\sin^2\varphi_1+a_2^2g_2^2(\varphi_2)+2a_1 a_2 g_1(\varphi_1) g_2(\varphi_2) \sin\varphi_1\sin\varphi_2]$.  Equation~\eqref{two_volkov_Bessel_n} carries a TAM value $j_{n_1,n_2}=l+n_1+n_2+s$. The OAM of the vortex electron is modified by both the LP and CP laser photons. \\

The probability density $\rho_{l,s,\text{two}}=\bar{\Psi}_{\text{two}}^{\text{vor}}\gamma^0 \Psi^{\text{vor}}_{\text{two}}$ is given by the following expression:
\begin{equation}
	\begin{aligned}
		\rho_{l,s,\text{two}} &\approx \sum_{n_1, n_2, n_1', n_2'} J_{n_1}(\tilde{\alpha}_{1, p_\perp}) J_{n_1'}(\tilde{\alpha}_{1, p_\perp}) J_{n_2}(\tilde{\alpha}_{2, p_\perp}) J_{n_2'}(\tilde{\alpha}_{2, p_\perp}) \\
		&\qquad \times \Bigl\{ \cos\Bigl[\Delta n_1 (\phi_{\bm{r}} + \tfrac{\pi}{2}) + \Delta n_2 (\tilde{\phi}_{\bm{r}} + \tfrac{\pi}{2})\Bigr] \\
		&\qquad \qquad \times \Bigl[ (1 - \tfrac{\Delta}{2}) J_{l+n_1'+n_2'} J_{l+n_1+n_2} \\
		&\qquad \quad + \tfrac{\Delta}{2} J_{l+n_1'+n_2'+2s} J_{l+n_1+n_2+2s} \Bigr] \Bigr\},
	\end{aligned}\label{two_rho_simp}
\end{equation}
where we have introduced $\tilde{\phi}_{\bm{r}}=\phi_{\bm{r}}-\varphi_2$ and defined $\Delta n_1=n_1-n'_1$ and $\Delta n_2=n_2-n'_2$. If we set $\tilde{\alpha}_{1,p_\perp}=0$ ($\tilde{\alpha}_{2,p_\perp}=0$) and  $n_1,n_1'=0$ ($n_2,n_2'=0$)  in Eq.~\eqref{two_rho_simp}, we recover the result for the CP (LP) laser pulse. Therefore, Eq.~\eqref{two_rho_simp} can be used to obtain the probability distribution of the vortex electron in the presence of individual LP and CP pulses as well as the combined two-mode laser pulse.\\ 

\section{Results and Discussion}

\begin{figure}[H]
	\begin{center}
	\setlength{\abovecaptionskip}{-0.0cm}
	\includegraphics[width=.95\linewidth]{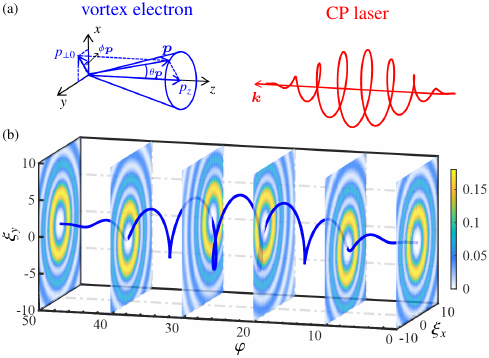}
	\begin{picture}(300,0)             
		\end{picture}
	\end{center}
	\caption{(a) Head-on collision of a vortex electron with a CP laser. (b) Propagation of the vortex electron in the CP laser pulse. Blue line: Classical trajectory of the point-charge electron propagating in a CP laser pulse. The probability density distributions of the vortex electron in the transverse ($\xi_x, \xi_y$) plane at various laser phase $\varphi$ values in arbitrary units are illustrated. The definitions $\xi_x = p_{\perp0} r \cos\phi_{\bm{r}}$ and $\xi_y = p_{\perp0} r \sin\phi_{\bm{r}}$ are employed. The momentum vectors $\bm{k}$ and $\bm{p}$ correspond to the laser photon and the plane-wave component of the vortex electron, respectively. The transverse momentum is fixed by $p_{\perp0}$, while $\theta_{\bm{p}}$ and $\phi_{\bm{p}}$ are the polar and azimuth angles of $\bm{p}$.}
	\label{fig_1}
\end{figure}

To study the propagation of vortex electrons in external laser fields, one can assume relatively low laser intensity ($a_0\ll 1$) and electron energy ($\varepsilon\sim$ MeV). This allows one to safely assume that the radiation is not present during the collision of a vortex electron with an external laser pulse. To characterize the motion of the vortex electron in the laser pulse, we calculate the probability density of finding an electron in the beam profile at a certain instant or laser phase $\varphi$.  
Upon collision with strong lasers [see Fig.~\ref{fig_1} (a)], the central point within the transverse plane of the vortex electron beam aligns with the trajectory of a classical point charge [depicted by the blue line in Fig.~\ref{fig_1} (b)]. The ring-shaped distribution that envelops this trajectory, as illustrated by the Bessel rings depicted in Fig.~\ref{fig_1}~(b), serves as a manifestation of the quantum broadening due to the vortex wave packet \cite{Karlovets:2012eu}.  Due to the adiabatic nature of the interaction between the laser field and the vortex electron, the vortex electron will resume its original vortex shape after exiting the laser field. Thus, we expect the Bessel ring in the transverse plane to return to its original position. In the following, we investigate these effects in more detail using numerical calculations. First, we consider the propagation of a vortex eigenmode electron and a superposition state electron in a CP laser pulse. Subsequently, we examine the propagation of a vortex electron in a combined laser field consisting of LP and CP laser pulses. Finally, we discuss the experimental feasibility of observing our results.

\subsection{Motion of the vortex electron in the CP laser pulse}
\begin{figure}[H]
	\begin{center}
	\setlength{\abovecaptionskip}{-0.0cm}
	\includegraphics[width=.9\linewidth]{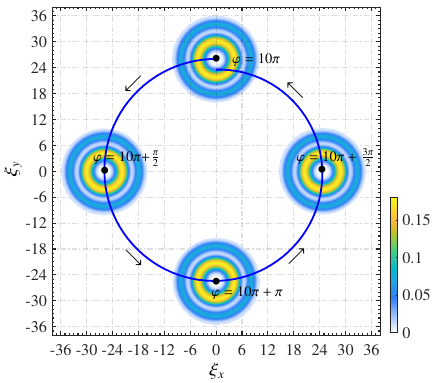}
	\begin{picture}(300,0)
		\end{picture}
	\end{center}
	\caption{Probability density distribution given by Eq.~\eqref{rho_simp} of the vortex electron in the transverse  $(\xi_x,\xi_y)$ plane at various laser phases in arbitrary units. Dots indicate the center of the probability density distribution of the vortex electron. The blue line represents the classical trajectory of an electron point charge in corresponding CP laser pulse.}
	\label{fig_2}
\end{figure}

In Fig.~\ref{fig_2}, the probability density distributions of the vortex electron in a CP laser pulse for different laser phase values are given.  The electron is assumed to be in a vortex eigenmode characterized by an OAM value of $l = 3$, an energy of $\varepsilon = 1.33$ MeV, and an absolute value of transverse momentum of $p_{\perp0}=868.3$ keV, corresponding to a polar angle of $\theta_{\bm{p}} = \frac{\pi}{4}$. The laser-pulse-shape function is defined as $g(\varphi) = \sin^2\left(\frac{\varphi}{2N_{\text{cycle}}}\right)$, where the number of laser cycles is set to $N_{\text{cycle}} = 10$. This corresponds to a pulse duration $\tau$ of approximately 26.7 fs for a laser frequency of $\omega = 1.55$ eV. For the chosen pulse-shape function, the dynamic integrals are found to be well approximated by the SVEA for $N_{\text{cycle}} \geq 6$. Therefore, our choice of parameters adheres to the SVEA, thereby validating its applicability in Eq.~\eqref{SVEA}. For relatively weak laser $\tilde{\delta}\ll 1$, we set $a_0=2\times10^{-4}$ and use Eq.~\eqref{rho_simp} to obtain the results. The numerical results show that the contributing values of $n$ range from $-20$ to $20$. The summation over electron spin $s$ is performed such that our results correspond to the unpolarized vortex electron. We present the probability distributions in the transverse $(\xi_x,\xi_y)$ plane for different laser phase values $\varphi=10\pi, 10\pi+\pi/2, 10\pi+\pi$, and $10\pi+3\pi/2$ in Fig.~\ref{fig_2}. \\

We observe that the center of the Bessel ring at each moment follows the classical motion of the point-charge electron, whose trajectory in the transverse plane can be expressed as 
\begin{equation}
\bm{r}_\perp(\varphi)=-\frac{e}{k\cdot p}\int^\varphi d\varphi' \bm{A}_\perp(\varphi')=\frac{m_e a_0}{k\cdot p} g(\varphi)\begin{pmatrix} -\sin\varphi \\ \cos\varphi \end{pmatrix}\,.
\label{class_trajectory}
\end{equation}
Due to the $\varphi$ dependence of the pulse-shape function $g(\varphi)$, the radius of the spiral trajectory varies for increasing laser phase $\varphi$.  The center of the probability distribution of the vortex electron starts at $(0,0)$ and gradually undergoes a spiral motion due to the CP laser pulse. In the monochromatic limit, where $g(\varphi)=1$, the electron is expected to follow a circular trajectory with a fixed radius \cite{Bu:2023fdj}.  The maximum distance from the center $(0,0)$, given as $\vert\tilde{\bm{r}}_\perp(10\pi)\vert = \frac{p_\perp m_e}{k \cdot p} a_0 \approx 26$ in the $(\xi_x,\xi_y)$ plane, corresponds to approximately 0.037 nm. 
To observe a considerable deviation from the center in the transverse plane, the maximum radius of the classical trajectory should be much larger than the beam waist of the vortex electron, $\vert\bm{r}_\perp\vert_{\text{max}}\gg\frac{\vert l\vert}{p_{\perp0}}$. Consequently, we obtain
\begin{equation}
\frac{a_0}{\omega}\gg\frac{\varepsilon+p_z}{p_{\perp0}m_e}\vert l\vert\,,
\label{a_0_condition}
\end{equation}
where, for the electron and laser pulse parameters considered in Fig~\ref{fig_2}, the condition reads, $a_0\gg2.3\times10^{-5}$. Therefore, to achieve a significant transverse displacement of the vortex beam, one needs an optical laser ($\omega = 1.55 \, \text{eV}$) with intensity $I_0 \gg 2.3 \times 10^9 \, \text{W}/\text{cm}^2$. The deviation of the vortex beam from the center increases with higher laser intensity because the radius of the trajectory in the transverse plane is proportional to the laser parameter $a_0 \sim \sqrt{I_0}$, as seen in Eq.~\eqref{class_trajectory}. \\

\begin{figure}[H]
	\begin{center}
	\setlength{\abovecaptionskip}{-0.0cm}
	\includegraphics[width=.9\linewidth]{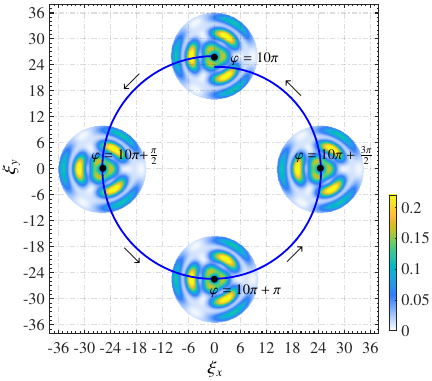}
	\begin{picture}(300,0)         
		\end{picture}
	\end{center}
	\caption{Probability density distribution given by Eq.~\eqref{sup_rho} of the vortex electron in the transverse  $(\xi_x,\xi_y)$ plane at various laser phases in arbitrary units. The electron is in a superposition state with OAM values $l_1=1$ and $l_2=4$. Dots indicate the center of the probability density distribution of the vortex electron. The blue line represents the classical trajectory of an electron point charge in the corresponding CP laser pulse.}
	\label{fig_3}
\end{figure}

Figure~\ref{fig_3} presents the probability density of the vortex electron propagating in a CP laser pulse, where the electron is in a superposition state with OAM values $l_1=1$ and $l_2=4$ and the relative phase is set to zero, ${\delta}^\ast=0$. We observe a ``flower like'' structure with petals characterized by $\Delta l=\vert l_1 - l_2\vert=3$, which is the same as that of a free-electron beam in the same vortex superposition state. During the collision, the center of the vortex beam follows a classical trajectory, similar to the motion of the vortex eigenmode depicted in Fig.~\ref{fig_2}, while the flower like structure remains unchanged. Additionally, no rotation about the beam's central axis is observed, confirming that the CP laser only causes orbital rotation of the vortex wave packet without disturbing its overall transverse structure. After the interaction, the center of the vortex electron beam returns to its original position in the transverse plane, coinciding with the position it had before interacting with the laser. \\

\subsection{Motion of the vortex electron in the two-mode laser pulse}

To achieve more diverse control over the behavior of vortex electrons, one could consider the combined effect of more than one laser. For instance, an LP laser, after frequency doubling, can be transformed into a CP laser by passing through an 800-nm 1/4-wave plate, which functions as a 400-nm 1/2-wave plate. This enables the interaction of vortex electrons with a two-mode laser field, allowing for manipulation of the vortex electrons through both LP and CP polarizations with different frequencies. As discussed in the previous section, such a setup is achievable in current experiments for vortex electrons with energies around MeV and two-mode optical lasers with $a_0$ values around $\sim10^{-4}$, enabling observable manipulation in the transverse plane.\\

\begin{figure}[H]
	\begin{center}
		\setlength{\abovecaptionskip}{-0.0cm}
		\includegraphics[width=.55\linewidth]{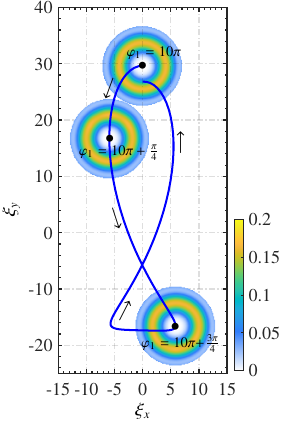}
		\begin{picture}(300,0)               
		\end{picture}
	\end{center}
	\caption{Probability density distribution given by Eq.~\eqref{two_rho_simp} of the vortex electron in the transverse  $(\xi_x,\xi_y)$ plane at various laser phases in arbitrary units. Dots indicate the center of the probability density distribution of the vortex electron. The blue line represents the classical trajectory of an electron point charge in the corresponding two-mode laser pulse. }
	\label{fig_x}
\end{figure}

Figure~\ref{fig_x} shows the results for a vortex electron interacting with the two-mode laser field given by Eq.~\eqref{two_field}, which is a combination of an LP laser and a CP laser pulse. The results are obtained using Eq.~\eqref{two_rho_simp}. The laser parameters are chosen as $a_{1,0} = 1 \times 10^{-4}$ and $a_{2,0} = 5 \times 10^{-5}$. The pulse-shape functions are defined as $g_1(\varphi_1) = \sin^2\left(\frac{\varphi_1}{2N_{\text{cycle}}}\right)$ and $g_2(\varphi_2) = \sin^2\left(\frac{\varphi_2}{2\nu N_{\text{cycle}}}\right)$, with $\nu = 2$. The polar angle is set to $\theta_{\bm{p}} = \frac{5}{12}\pi$, and all other parameters are the same as in Fig.~\ref{fig_2}. The probability distributions are presented at laser phase values $\varphi_1 = 10\pi$, $10\pi + \frac{\pi}{4}$, and $10\pi + \frac{3\pi}{4}$.  The classical trajectory of the point-charge electron inside the two-mode laser field is given by, 
\begin{equation}
	\bm{r}_\perp=\frac{m_e a_{1,0}}{k_1\cdot p} g_1(\varphi_1)\begin{pmatrix} 0 \\ \cos\varphi_1  \end{pmatrix}+\frac{m_e a_{2,0}}{k_2\cdot p} g_2(\varphi_2)\begin{pmatrix} -\sin\varphi_2 \\ \cos\varphi_2   \end{pmatrix}\,.
	\label{class_trajectory}
\end{equation}
The LP laser shifts the electron vertically, while the CP laser induces a rotation or spiral motion. The horizontal deviation is smaller due to the classical trajectory's smaller value in the corresponding direction, given by $\frac{m_e a_{2,0}}{k_2\cdot p} < \frac{m_e a_{1,0}}{k_1\cdot p}$. The vertical deviation shows significant asymmetry due to the pulse shape and the differing contributions of the CP and LP pulses. This enables more flexible control of the electron's motion within two-mode lasers compared to using LP or CP lasers alone. The rich variety of classical motions facilitated by two-mode lasers with different polarizations and frequencies allows for diverse manipulation of vortex electrons at subnanometer and femtosecond scales. These novel Volkov-Bessel modes expand the range of achievable vortex modes, potentially stimulating new applications in using electron waves as probes for special materials \cite{VERBEECK2014190}.\\

\subsection{Discussion}
The predicted motion of the vortex electron in the laser field can be detected via the pump-probe method \cite{Hayrapetyan:2014faa,RevModPhys.81.163,GRUMSTRUP201530}. In this method, a pump laser, which serves as the external laser field in our work, collides with the electron beam and induces transverse motion of the vortex electrons. The probe laser, which is weak enough not to perturb the electron beam, can then be used to reveal the spatiotemporal structure of the electron beam. \\

For the vortex electron created in the paraxial regime, we set the polar angle to $\theta_{\bm{p}} = 0.02$ mrad with energy $\varepsilon = 1.33$ MeV and obtain an estimate for the pump laser that can significantly influence the dynamics of the vortex electron: $a_0 \gg 9.5 \times 10^{-4}$ or $I_0 \gg 4 \times 10^{12} \, \text{W/cm}^2$ (for $\omega = 1.55 \text{eV}$). 
Moreover, since $\frac{a_0}{\omega} \propto \sqrt{I_0} \lambda_0^2$, one can infer from the condition given by Eq.~\eqref{a_0_condition} that increasing the laser intensity $I_0$ and/or employing lasers with a larger wavelength $\lambda_0$ can induce a more significant transverse deviation of the vortex-electron-beam center.\\

\section{Conclusion}
We investigated the dynamics of relativistic vortex electrons during their head-on collisions with plane-wave laser pulses. Our findings reveal that the beam center of the vortex electron follows the classical trajectory of a point-charge electron propagating in the same laser field. Specifically, the center of the vortex electron beam undergoes rotation in CP laser pulses, experiences a lateral shift in LP laser pulses, and exhibits a combined twisted spiral pattern in the superposition of LP and CP laser fields. Importantly, these dynamics occur while maintaining the transverse structure of the vortex eigenstate and vortex superposition states. These effects can be detected using pump-probe experiments with the help of electron microscopy.\\

Our results demonstrate the potential for manipulating electron vortex beams using various laser modes.  Our work provides a foundation for future experimental and theoretical studies, serving as a benchmark reference for investigations concerning the manipulation of vortex electron beams using more realistic laser or other types of external fields. As a next step, one can consider a rich class of lasers, including the effects of chirp, focusing, and OAM. Additionally, investigating other electron vortex modes, such as Laguerre-Gaussian modes, could provide further insights. Finally, for higher electron energies and higher laser intensities, it would be valuable to investigate the radiation emitted by the vortex electrons.\\

\begin{flushright}
	
\end{flushright}

{\it Acknowledgments} This work is supported by the National Natural Science Foundation of China (Grants No. U2267204, No. 12425510, and No. 1240050217), the Foundation of Science and Technology on Plasma Physics Laboratory (Grant No. JCKYS2021212008), the Natural Science Basic Research Program of Shaanxi (Grants No. 2023-JC-QN-0091 and No. 2024JC-YBQN-0042), and the Shaanxi Fundamental Science Research Project for Mathematics and Physics (Grants No. 22JSY014, No. 22JSQ019, and No. 23JSQ006).\\


\appendix
\section{Current Density for a Vortex Electron in a CP Laser Pulse}
The four-current density for a vortex electron in a CP laser pulse is defined as $\mathcal{J}^\mu=(\rho_{l,s},\bm{j}_{l,s})=\bar{\Psi}^{\text{vor}}\gamma^\mu \Psi^{\text{vor}}$, and the probability density $\rho_{l,s}$ is given in the main text in Eq.~\eqref{rho}. The components for the current density $\bm{j}_{l,s}$ are given by
\begin{equation}
	\begin{aligned}
{j}^{(x)}_{l,s}=&\sum_{n,n^\prime} J_{n^\prime}( \tilde{\alpha}_{p_\perp}  )J_n(\tilde{\alpha}_{p_\perp}) \Biggl(
\cos \bigl[ \Delta n \bigl(\tilde{\phi}_{\bm{r}}+\frac{\pi}{2}\bigr) \bigr] \cos\varphi \\
&\times(-\tilde{\delta})\Biggl[\rho_{lsnn^\prime}(\xi)+\frac{p_z}{\varepsilon}J_{l+n^\prime}J_{l+n}\Biggr] \\
&-\sin \bigl[ 2s \Delta n \bigl(\tilde{\phi}_{\bm{r}}+\frac{\pi}{2}\bigr) + {\phi}_{\bm{r}} \bigr] 
\frac{ p_\perp}{\varepsilon}J_{l+n^\prime}J_{l+n+2s} \Biggr)\,,
	\end{aligned}\label{current_x}
\end{equation}		
\begin{equation}
	\begin{aligned}
		{j}^{(y)}_{l,s}=&\sum_{n,n^\prime} J_{n^\prime}( \tilde{\alpha}_{p_\perp}  )J_n(\tilde{\alpha}_{p_\perp}) \Biggl(
		\cos \bigl[ \Delta n \bigl(\tilde{\phi}_{\bm{r}}+\frac{\pi}{2}\bigr) \bigr] \sin\varphi \\
		&\times(-\tilde{\delta})\Biggl[\rho_{lsnn^\prime}(\xi)+\frac{p_z}{\varepsilon}J_{l+n^\prime}J_{l+n}\Biggr] \\
		&+\cos \bigl[ 2s \Delta n \bigl(\tilde{\phi}_{\bm{r}}+\frac{\pi}{2}\bigr) + {\phi}_{\bm{r}} \bigr] 
		\frac{p_\perp}{\varepsilon}J_{l+n^\prime}J_{l+n+2s} \Biggr)\,,
	\end{aligned}\label{current_y}
\end{equation}
and
\begin{equation}
	\begin{aligned}
		{j}^{(z)}_{l,s}=&\sum_{n,n^\prime} J_{n^\prime}( \tilde{\alpha}_{p_\perp} )J_n(\tilde{\alpha}_{p_\perp}) \Biggl(
		\cos \bigl[ \Delta n \bigl(\tilde{\phi}_{\bm{r}}+\frac{\pi}{2}\bigr) \bigr] \\
		&\times\Biggl[-\frac{\tilde{\delta}^2}{2}\rho_{lsnn^\prime}(\xi)+(1-\frac{\tilde{\delta}^2}{2})\frac{p_z}{\varepsilon}J_{l+n^\prime}J_{l+n}\Biggr] \\
		&-\sin \bigl[ 2s \Delta n \bigl(\tilde{\phi}_{\bm{r}}+\frac{\pi}{2}\bigr) + \tilde{\phi}_{\bm{r}} \bigr] 
		\frac{\tilde{\delta} \, p_\perp}{\varepsilon}J_{l+n^\prime}J_{l+n+2s} \Biggr)\,.
	\end{aligned}\label{current_z}
\end{equation}
The transverse current $\bm{j}^{\perp}_{l,s}=({j}^{(x)}_{l,s},{j}^{(y)}_{l,s})$ can be written in the polar coordinates $\bm{j}^{\perp}_{l,s}=({j}^{(r)}_{l,s},{j}^{(\phi)}_{l,s})$ using the following relation:
\begin{equation}
\begin{aligned}
{j}^{(r)}_{l,s}&={j}^{(x)}_{l,s}\cos\phi_{\bm{r}} + {j}^{(y)}_{l,s} \sin\phi_{\bm{r}}\,, \\
{j}^{(\phi)}_{l,s}&=-{j}^{(x)}_{l,s}\sin\phi_{\bm{r}} + {j}^{(y)}_{l,s} \cos\phi_{\bm{r}}\,. \\
\end{aligned}
\end{equation}
The transverse currents in polar coordinates read
\begin{equation}
	\begin{aligned}
		{j}^{(r)}_{l,s}=&\sum_{n,n^\prime} J_{n^\prime}( \tilde{\alpha}_{p_\perp}  )J_n(\tilde{\alpha}_{p_\perp}) \Biggl(
		\cos \bigl[ \Delta n \bigl(\tilde{\phi}_{\bm{r}}+\frac{\pi}{2}\bigr) \bigr] \cos\tilde{\phi}_{\bm{r}} \\
		&\times(-\tilde{\delta})\Biggl[\rho_{lsnn^\prime}(\xi)+\frac{p_z}{\varepsilon}J_{l+n^\prime}J_{l+n}\Biggr] \\
		&-\sin \bigl[ 2s \Delta n \bigl(\tilde{\phi}_{\bm{r}}+\frac{\pi}{2}\bigr)  \bigr] 
		\frac{p_\perp}{\varepsilon}J_{l+n^\prime}J_{l+n+2s} \Biggr)\,,
	\end{aligned}\label{current_r}
\end{equation}
and
\begin{equation}
	\begin{aligned}
		{j}^{(\phi)}_{l,s}=&\sum_{n,n^\prime} J_{n^\prime}( \tilde{\alpha}_{p_\perp}  )J_n(\tilde{\alpha}_{p_\perp}) \Biggl(
		\cos \bigl[ \Delta n \bigl(\tilde{\phi}_{\bm{r}}+\frac{\pi}{2}\bigr) \bigr] (-\sin\tilde{\phi}_{\bm{r}}) \\
		&\times(-\tilde{\delta})\Biggl[\rho_{lsnn^\prime}(\xi)+\frac{p_z}{\varepsilon}J_{l+n^\prime}J_{l+n}\Biggr] \\
		&+\cos \bigl[ 2s \Delta n \bigl(\tilde{\phi}_{\bm{r}}+\frac{\pi}{2}\bigr)  \bigr] 
		\frac{p_\perp}{\varepsilon}J_{l+n^\prime}J_{l+n+2s} \Biggr)\,.
	\end{aligned}\label{current_phi}
\end{equation}
The transverse currents also possess a ``mirror-reflection" symmetry, as can be shown by the invariance of the current densities under the replacement $\tilde{\phi}_{\bm{r}} \rightarrow \tilde{\phi}_{\bm{r}} - \pi$. Since the azimuth angle variable $\tilde{\phi}_{\bm{r}}$ depends on the laser phase ($\tilde{\phi}_{\bm{r}} = \phi_{\bm{r}} - \varphi$), the symmetry axis actually varies with the laser phase. This is expected because the vortex electron follows the classical motion of a point-charge electron, such that in a CP laser pulse there is a spiral rotation in the transverse plane, which also rotates the symmetry axis of the current density.\\

In the case of a relatively weak laser, we assume $\tilde{\delta}\rightarrow0$ and obtain the following approximate results for the current densities: 
\begin{equation}
	\begin{aligned}
		{j}^{(z)}_{l,s}\approx & \sum_{n,n^\prime}J_{n^\prime}( \tilde{\alpha}_{p_\perp} )J_n(\tilde{\alpha}_{p_\perp}) \\
		&\times \frac{p_z}{\varepsilon} \cos \bigl[ \Delta n \bigl(\tilde{\phi}_{\bm{r}}+\frac{\pi}{2}\bigr) \bigr]J_{l+n^\prime}J_{l+n} \,,
	\end{aligned}\label{simp_current_z}
\end{equation}
\begin{equation}
	\begin{aligned}
		{j}^{(r)}_{l,s}\approx&-\sum_{n,n^\prime} J_{n^\prime}( \tilde{\alpha}_{p_\perp}  )J_n(\tilde{\alpha}_{p_\perp})\\
		&\times  \frac{p_\perp}{\varepsilon} \sin \bigl[ 2s \Delta n \bigl(\tilde{\phi}_{\bm{r}}+\frac{\pi}{2}\bigr)  \bigr] 
		J_{l+n^\prime}J_{l+n+2s},
	\end{aligned}\label{simp_current_r}
\end{equation}
and 
\begin{equation}
	\begin{aligned}
		{j}^{(\phi)}_{l,s}\approx &\sum_{n,n^\prime} J_{n^\prime}( \tilde{\alpha}_{p_\perp}  )J_n(\tilde{\alpha}_{p_\perp}) \\
		&\times  \frac{p_\perp}{\varepsilon} \cos \bigl[ 2s \Delta n \bigl(\tilde{\phi}_{\bm{r}}+\frac{\pi}{2}\bigr)  \bigr] J_{l+n^\prime}J_{l+n+2s}\,.
	\end{aligned}\label{simp_current_phi}
\end{equation}
If we make the replacements $\tilde{\phi}_{\bm{r}} \rightarrow \phi_{\bm{r}}$ and $\tilde{\alpha}_{p_\perp} = \alpha_{p_\perp} g(\varphi) \rightarrow \tilde{\alpha}_{p_\perp} = \alpha_{p_\perp} g(\varphi) \cos \varphi$, we obtain the results for the transverse current density in polar coordinates for LP laser pulses.\\

In the weak laser limit, the $z$ component of the current density becomes spin independent, as seen by comparing Eq.~\eqref{current_z} to Eq.~\eqref{simp_current_z}. In contrast, both the $r$ and $\phi$ components are spin-dependent and take very small values in the paraxial limit where $\frac{p_\perp}{\varepsilon} \ll 1$, as shown in Eqs.~\eqref{simp_current_r} and \eqref{simp_current_phi}.\\

\section{Complete Expressions for Probability Density}

\subsection{Vortex superposition state electron in the CP laser pulse}
The complete expression for the probability density of the vortex superposition electron in the CP laser pulse reads
\begin{equation}
	\begin{aligned}
		\rho_{l_1l_2s} &= \sum_{n,n'} J_{n'}(\tilde{\alpha}_{p_\perp}) J_{n}(\tilde{\alpha}_{p_\perp}) \frac{1}{2} \Biggl\{
		(1 + \frac{\tilde{\delta}^2}{2}) \cos[\Delta n(\tilde{\phi}_{\bm{r}} - \frac{\pi}{2})] \\
		&\quad \times \Biggl[(1 - \frac{\Delta}{2}) (J_{l_1+n'} J_{l_1+n} + J_{l_2+n'} J_{l_2+n}) \\
		&\quad + \frac{\Delta}{2} (J_{l_1+n'+2s} J_{l_1+n+2s} + J_{l_2+n'+2s} J_{l_2+n+2s})
		\Biggr] \\
		&+ 2 \cos[\Delta n(\tilde{\phi}_{\bm{r}} + \frac{\pi}{2}) + \Delta l \phi_{\bm{r}} + \delta^\ast] \Biggl[
		(1 - \frac{\Delta}{2}) J_{l_1+n'} J_{l_2+n} \\
		&\quad + \frac{\Delta}{2} J_{l_1+n'+2s} J_{l_2+n+2s}
		\Biggr] \\
		&+ \frac{\tilde{\delta}^2}{2} \frac{p_z}{\varepsilon} \Biggl[
		\cos[\Delta n(\tilde{\phi}_{\bm{r}} - \frac{\pi}{2})] (J_{l_1+n'} J_{l_1+n} + J_{l_2+n'} J_{l_2+n}) \\
		&\quad + 2 \cos[\Delta n(\tilde{\phi}_{\bm{r}} + \frac{\pi}{2}) + \Delta l \phi_{\bm{r}} + \delta^\ast] \Biggl[
		(1 - \frac{\Delta}{2}) J_{l_1+n'} J_{l_2+n} \\
		&\quad + \frac{\Delta}{2} J_{l_1+n'+2s} J_{l_2+n+2s}
		\Biggr]
		\Biggr] \\
		&+ \tilde{\delta} \frac{p_\perp}{\varepsilon} \Biggl[
		\sin[2s\Delta n (\tilde{\phi}_{\bm{r}} + \frac{\pi}{2}) + \tilde{\phi}_{\bm{r}}] \\
		&\quad \times (J_{l_1+n'} J_{l_1+n+2s} + J_{l_2+n'} J_{l_2+n+2s}) \\
		&\quad + \sin[2s\Delta n (\tilde{\phi}_{\bm{r}} + \frac{\pi}{2}) + 2s\Delta l \phi_{\bm{r}} + \tilde{\phi}_{\bm{r}} + {\delta}^\ast] \\
		&\quad \times J_{l_1+n'} J_{l_2+n+2s} \\
		&\quad + \sin[2s\Delta n (\tilde{\phi}_{\bm{r}} + \frac{\pi}{2}) - 2s\Delta l \phi_{\bm{r}} + \tilde{\phi}_{\bm{r}} - {\delta}^\ast] \\
		&\quad \times J_{l_2+n'} J_{l_1+n+2s}
		\Biggr]
		\Biggr\}.
	\end{aligned}
	\label{sup_rho_complete}
\end{equation}
By setting $\tilde{\delta}\rightarrow0$, we could obtain the simpler result in Eq.~\eqref{sup_rho} for relatively weak laser intensities. Moreover, if we set $l_1,l_2=l$ and $\delta^\ast=0$ in Eq.~\eqref{sup_rho_complete}, we recover the complete result for vortex eigenmode electron given by Eq.~\eqref{rho}. \\

\subsection{Vortex electron in the two-mode laser pulse}

The complete expression for the probability density of the vortex electron in the two-mode laser pulse reads,
\begin{equation}
	\begin{aligned}
		\rho_{l,s,\text{two}} &= \sum_{n_1,n_2,n_1^\prime,n_2^\prime} J_{n_1}(\tilde{\alpha}_{1,p_\perp}) J_{n_1^\prime}(\tilde{\alpha}_{1,p_\perp}) J_{n_2}(\tilde{\alpha}_{2,p_\perp}) J_{n_2^\prime}(\tilde{\alpha}_{2,p_\perp}) \bigr\{ \\
		&\times \cos\bigl[\Delta n_1 (\phi_{\bm{r}} + \frac{\pi}{2}) + \Delta n_2(\tilde{\phi}_{\bm{r}} + \frac{\pi}{2})\bigr] \bigl( (1 + \frac{\tilde{\delta}_1^2}{2}\sin^2\varphi_1 \\
		&\quad + \frac{\tilde{\delta}_2^2}{2} + \tilde{\delta}_1\tilde{\delta}_2 \sin\varphi_1\sin\varphi_2) \bigl[(1 - \frac{\Delta}{2}) J_{l+n_1^\prime+n_2^\prime} J_{l+n_1+n_2} \\
		&\quad + \frac{\Delta}{2} J_{l+n_1^\prime+n_2^\prime+2s} J_{l+n_1+n_2+2s}\bigr] + \bigl(\frac{\tilde{\delta}_1^2}{2}\sin^2\varphi_1 \\
		&\quad + \frac{\tilde{\delta}_2^2}{2} + \tilde{\delta}_1\tilde{\delta}_2 \sin\varphi_1\sin\varphi_2\bigr) \frac{p_z}{\varepsilon} J_{l+n_1^\prime+n_2^\prime} J_{l+n_1^\prime+n_2^\prime} \bigr)  \\
		&+ \sin\bigl\{2s\bigl[\Delta n_1 (\phi_{\bm{r}} + \frac{\pi}{2}) + \Delta n_2(\tilde{\phi}_{\bm{r}} + \frac{\pi}{2})\bigr] + \tilde{\phi}_{\bm{r}} \bigr\} \\
		&\times \tilde{\delta}_2 \frac{p_\perp}{\varepsilon} J_{l+n_1^\prime+n_2^\prime} J_{l+n_1+n_2+2s} + \cos\bigl\{2s\bigl[\Delta n_1 (\phi_{\bm{r}} + \frac{\pi}{2}) \\
		&\quad + \Delta n_2(\tilde{\phi}_{\bm{r}} + \frac{\pi}{2})\bigr] + {\phi}_{\bm{r}} \bigr\} \\
		&\times(-\tilde{\delta}_1\sin\varphi_1) \frac{p_\perp}{\varepsilon} J_{l+n_1^\prime+n_2^\prime} J_{l+n_1+n_2+2s} \bigr\}.
	\end{aligned}\label{comp_two_rho}
\end{equation}
By setting $\tilde{\delta}_1\rightarrow0$ and $\tilde{\delta}_2\rightarrow0$, we obtain the simplified result in Eq.~\eqref{two_rho_simp} for relatively weak laser intensities. The $\phi_{\bm{r}}$ dependence of the LP terms contrasts with the $\tilde{\phi}_{\bm{r}}$ dependence of the CP terms. Using Eq.~\eqref{comp_two_rho}, we can calculate the probability distribution of an electron in a superposition of LP and CP laser pulses for vortex electrons in both paraxial and nonparaxial limits across a wide range of laser intensities. One can also obtain the result for a LP (CP) laser pulse by setting $\tilde{\alpha}_{2,p_\perp}=0$ ($\tilde{\alpha}_{1,p_\perp}=0$) and $n_2,n'_2=0$ ($n_1,n'_1=0$). \\

\section{Comparing current densities for Volkov electrons and Volkov-Bessel electrons}
We compare the four-current densities of a vortex electron and a plane-wave electron propagating in a laser field. The plane-wave electron interacting with a laser field is described by the Volkov state, given by Eq.~\eqref{Volkov}. The corresponding four-current density $ j^\mu = \bar{\Psi}_p \gamma^\mu \Psi_p $ is expressed as \cite{Berestetskii1982quantum}
	\begin{equation}
		j^\mu = \frac{1}{\varepsilon} \left\{ p^\mu - e A^\mu(\varphi) + k^\mu \left[ \frac{e p \cdot A(\varphi)}{k \cdot p} - \frac{e^2 A^2(\varphi)}{2 k \cdot p} \right] \right\},
		\label{current_plane}
	\end{equation}
	where the position dependene of the four-current density is introduced by the laser field through the phase $\varphi = \omega t - k_z z$. Notably, there is no transverse coordinate dependency in Eq.~\eqref{current_plane}.\\
	
	In contrast, for the vortex electrons considered in this work, the four-current density depends not only on $\varphi$ but also on the transverse coordinates $\phi_{\bm{r}}$ and $r$ (through $\xi = p_\perp r$), as seen in Eqs.~\eqref{rho}, \eqref{current_x}, \eqref{current_y}, and \eqref{current_z}. This additional transverse dependence highlights the distinct structure of the vortex electrons compared to plane-wave electrons. Our numerical results in Sec. \uppercase\expandafter{\romannumeral3}  demonstrate that the transverse dependence of the probability density of the vortex electron manifests as the center of the Bessel rings following the classical trajectory of a point-charge electron.\\
	
	Moreover, the current density of plane electrons is spin-independent, whereas the current density of vortex electrons becomes spin-dependent when the intrinsic SOI parameter $\Delta$ is non-negligible. For relativistic vortex electrons, the current density depends on the electron spin $s$ in the nonparaxial regime, characterized by $\theta_{\bm{p}} \not\ll 1$. 
\bibliography{EVB_blib}

\end{document}